\documentclass[
]{ceurart}

\usepackage{listings}
\usepackage{longtable}
\usepackage{multirow}
\usepackage{graphicx}
\usepackage[export]{adjustbox}
\begin{document}

\copyrightyear{2022}
\copyrightclause{Copyright for this paper by its authors.
  Use permitted under Creative Commons License Attribution 4.0
  International (CC BY 4.0).}

\conference{The 10th edition of the PhD Symposium on FDIA, July 20, 2022, Lisbon, Portugal.}

\title{Text Information Retrieval in Tetun:\\
A Preliminary Study}


\author[1,2]{Gabriel de Jesus}[%
orcid=0000-0003-4392-2382,
email=gabriel.jesus@inesctec.pt
]
\address[1]{INESC TEC - Institute for Systems and Computer Engineering, Technology and Science, Portugal}
\address[2]{FEUP - Faculty of Engineering, University of Porto, Portugal}


\begin{abstract}
  Tetun is one of Timor-Leste’s official languages alongside Portuguese. It is a low-resource language with over 932,400 speakers that started developing when Timor-Leste restored its independence in 2002. The media mainly uses Tetun, and more than ten national online newspapers actively broadcast news in Tetun every day. However, since information retrieval-based solutions for Tetun do not exist, finding Tetun information on the internet is challenging. This work aims to investigate and develop solutions that can enable the application of information retrieval techniques to develop search solutions for Tetun. We present a preliminary result of an experiment conducted on the task of ad-hoc retrieval in Tetun.
\end{abstract}


\maketitle

\section{Introduction and Motivation}

Information retrieval deals with finding documents of unstructured-nature (usually text) that satisfies an information need from within a large collection \cite{manning-et-al-09}. Users of the retrieval system tend to search for information on a topic of their interest and express it in many different ways, often in very different words using natural language text \cite{croft-et-al-15}. They usually formulate their information needs in a set of words in textual-based queries and expect to get the documents relevant to their information needs. One important information retrieval application is keyword-based web search, as showcased by the Google search engine. Since specific information retrieval-based approaches for Tetun do not exist, it becomes more challenging to find the documents that satisfy information needs written in Tetun.

Tetun is the language spoken in Timor-Leste. There are two major varieties of Tetun: Tetun Dili (referred to as Tetun) and Tetun Terik \cite{klinken-et-al-02}. Tetun is the most spoken language in the country, and Tetun Terik is one of the Timor-Leste dialects. Most of the Tetun verb, noun, and adjective are Portuguese loanwords \cite{klinken-hajek-18, greksakova2018tetun}. Klinken and Hajek \cite{klinken-hajek-18} studied a selection of seven articles from different newspapers in 2009 and stated that an average of 32\% words are Portuguese loans, and Greksáková \cite{greksakova2018tetun} reported 35\% of Portuguese loanwords in analyzing 73,892 words from the interview transcripts corpus. As per the census 2015 report \cite{census-tl-15}, the population of Timor-Leste was 1.17 million and the proportion of the Tetun speakers was 79.04\%, including Portuguese (2.56\%), Indonesian (2.02\%), and English (1.04\%). Tetun became a dominant language in public life when the Government of Timor-Leste designated it as one of the Timor-Leste's official languages in 2002. Then, the Government established the National Institute of Linguistics (INL) \cite{dl-rdtl-no1-04} and produced the Standard Orthography of Tetun Language (Tetun INL) \cite{inl-04}.

\begin{wrapfigure}{r}{0.67\textwidth}
\includegraphics[width=\linewidth]{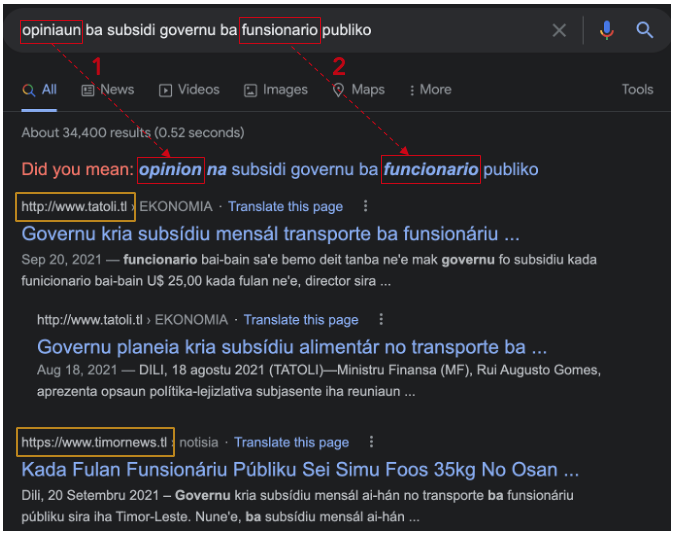} 
\caption{Example of searching using Tetun in Google Search.}
\label{fig:figure1}
\end{wrapfigure}

Although Tetun INL was already in place, the Australian linguist Catharina v. Klinken and her colleagues in the Tetun Department of the Dili Institute of Technology (DIT) disputed the orthography and established their standard for Tetun (Tetun DIT) \cite{klinken-et-al-02}. There are some differences between Tetun INL and Tetun DIT in writing structures, such as Tetun INL uses “ñ” and “ll”, but Tetun DIT uses “nh” and “lh”, e.g., millaun (Tetun INL), milhaun (Tetun DIT). Consequently, there is still a wide variation on the use of Tetun in official writing and media. Moreover, Tetun is also influenced by Portuguese, Indonesian, and English spoken in Timor-Leste. 
The query in \autoref{fig:figure1} illustrates a search performed by one of the Timor News visitors and captured by the platform – [ opiniaun ba \textit{subsidi} governu ba \textit{funsionario publiko} ] – ``subsidi'' is Indonesian (Tetun: subsídiu) and ``funsionario publiko'' are mixed of Portuguese and Tetun (Tetun: funsionáriu públiku). We copied this query to the Google search engine and executed it to observe the retrieved results. As shown in note~1, Google suggested the \textit{opiniaun} word in Tetun to English and the \textit{funsionario} in note~2 to Portuguese. This is the evidence that motivates us to contribute to the development of information retrieval techniques for Tetun.

To tackle the aforementioned problem, this work aims to focus on developing solutions for text information retrieval tasks for Tetun, including the development of Tetun corpus. Based on the Google indicator, there are around 4.8 million documents available on the internet from the sites hosted under ``.tl'' top-level domain. Therefore, we will study them in detail to identify the text documents that are potentially relevant to our work. Considering that online news has played a crucial role in promoting Tetun in the last five years, from 2017 to 2022, as also indicated in the documents retrieved for the query in \autoref{fig:figure1}, we will use Timor News\footnote{\url{https://www.timornews.tl}} as our case study. Timor News will be used to test the search solutions prototype for Tetun to evaluate further the algorithms' performance. We conducted an initial indexing and retrieval experiment using documents from Timor News, where our preliminary result is in Section \ref{sec:preliminary_study_result}. Timor New is an online news agency based in Dili, Timor-Leste, that only broadcasts national and international news in Tetun, and as it is founded by the author of this paper, we have full access to the platform (documents, search queries, etc.).  

The remainder of this paper is organized as follows. Section \ref{sec:related_work} presents related work. We describe our methodology in Section \ref{sec:methodology}. Then, Sections \ref{sec:research_question} and \ref{sec:preliminary_study_result} present research question and preliminary study results. Finally, Section \ref{sec:conclusion_future_work} summarizes our conclusion and planned future work.

\section{Related Work}
\label{sec:related_work}

Low-resource languages can be understood as languages that are less studied, less computerized, low-density \cite{Magueresse2020LowresourceLA}, that lack text corpora, or reduced accessibility \cite{Hoenen_Koc_Rahn_2020}. Developing information retrieval tools and techniques for a low-resource language requires the availability of corpora and test collections. Moreover, few algorithms support information retrieval for low-resource languages, and considering linguistic diversity, some language-specific algorithms, e.g., for text pre-processing and normalization, also need to be developed.

Different strategies to build datasets for low-resource languages have been studied. Artetxe et al.~\cite{artetxe-et-al-22} built the Basque corpus using an ad-hoc crawling approach — crawling 33 websites as the data sources — and stated that ad-hoc crawling could be an effective alternative to obtain high-quality data for low-resource languages. Issam and Mrini \citep{issam2021goudma} and Mbonu et al.~\cite{mbonu-et-al-22} built the datasets for text summarization tasks for Moroccan Darija and Igbo and made them publicly available. The first authors used GOUD.MA news website to build the GOUD.MA dataset, and the latter used the Anambra Broadcasting Service (ABS) website to build the IgboSum1500 dataset. Getao and Mariti \cite{getao-mariti-05} built the Kiswahili corpus employing a query-based crawling. The queries were manually created using bigram model and then crawled the documents for each queries. Hoenen et al.~\cite{Hoenen_Koc_Rahn_2020} developed a manual for web corpus crawling of low-resource languages with several data collection approaches. Moreover, the practical tasks of web corpus construction described by Schäfer and Bildhauer \cite{Schfer2013WebCC} can be used as the main reference. Some techniques for dataset construction are presented in \autoref{tab:datasets}.

\begin{longtable}{@{}lllll@{}}
\caption{Dataset Construction Techniques.}
\label{tab:datasets}\\
\toprule
Technique       & Data source         & \#Documents (or words) & Language                   \\
\midrule
Ad-hoc crawling & 33 websites         & 12.5M documents       & Basque               \\
Scraping & ABS website & 1.5k documents & Igbo 
\\
Scraping        & News websites       & 158k documents       & Moroccan Darija       \\
Query-based crawling & World Wide Web & 4.84M words & Kiswahili \\*
 \bottomrule
\end{longtable}

For the text processing tasks, Jurish et al.~\cite{jurish-et-al-13} proposed a Hidden Markov Model-based approach to segment automatically text documents into tokens and sentences. Ferilli \cite{ferilli-2021} presented a term-document frequency approach that automatically detects stopwords from a small amount of the corpora and stated that it was the most effective approach which outperformed the classic term frequency \cite{ferilli-et-al-2021, croft-et-al-15, yates-neto-11} and the normalized inverse document frequency of Lo et al.~\cite{Lo-et-al-05}. Trishala and Mamatha \cite{trishala-mamatha-2021} proposed a rule-based Kannada stemmer relying on an unsupervised approach using k-means algorithm, and Thangarasu and Inbarani \cite{mathan-inbarani-16} presented an analogy removal stemmer that automatically stem Tamil words from the text corpora. The state-of-art of query spelling corrections explained by Chang et al.~\cite{chang-deng-2020} are approaches applicable for low- and high-resource languages.

The TREC-style approach based on the Cranfield paradigm is used in developing test collections for low-resource languages. Chavula and Suleman \cite{chavula-suleman-21} built a test collection for Chichewa, Citumbuka, and Cinyanja, and the relevance assessment was conducted by the five external evaluators. Esmaili et al.~\cite{esmaili-et-al-13} built a Pewan test collection for Sorani Kurdish and Aleahmad et al.~\cite{aleahmad-et-al-09} created a Hamshahri test collection for Persian. The first test collection was judged by the team members (native Sorani speakers), and Persian students collaborated in the latter. Both test collections were built using news datasets. The details are in \autoref{tab:test-collection}.

\begin{longtable}{@{}llllll@{}}
\caption{Test Collection Details.}
\label{tab:test-collection}\\
\toprule
Name           & \#Documents & \#Queries & \#Run & \#Pool documents & \#Final Queries \\* \midrule
\endfirsthead
\endhead
C3\footnote{In the paper, the collection has no name assigned. We designated C3 to represent Chichewa, Citumbuka, and Cinyanja.} & 11,778 & 387 & 4 & 38,700 & 387 \\
Pewan & 115,340       & 42        & 5  & 2,500   & 22\footnote{For the reliability, queries with less than ten or more than 100 relevant documents were excluded.}              \\
Hamshahri      & 166,774       & 65  & 7  & 6,500   & 62\footnote{Three queries from the collection that have less than ten or more than 90 relevant documents were omitted.}              \\* \bottomrule
\end{longtable}

There are several existing search engines for low-resource languages, such as the works of Malumba et al.~\cite{malumba-et-al-15} and Kyeyune \cite{isixhosa-report-15} with search engines for Bantu languages, IsiZulu and IsiXhosa. Both search engines are developed using the same infrastructure — crawling the documents from the web and indexing and retrieving them using the Solr platform.

Regarding work for Tetun text, to the best of our knowledge, no one has conducted research in this area so far.

\section{Methodology}
\label{sec:methodology}

In this section, we present the proposed steps for our PhD work focused on text information retrieval in Tetun.

\paragraph{Literature Review}
Review the existing research on text information retrieval, specifically the works for low-resource languages, to understand the concepts, methods, and evidence that can be used to support our work. This step includes studying and comparing the methods for document collection and characterization and text pre-processing development.

\paragraph{Documents Collection and Characterization}
Build the collection of documents by collecting news articles and search queries from the Timor News database. To enrich the collections, we will develop a web crawler to collect text documents in Tetun from the sites hosted under Timor-Leste top-level domain ``.tl''. The Tumba!, a search engine that involves crawling sites hosted under the ``.pt'' top-level domain, including domains that contain a large set of Portuguese pages with incoming links from web pages hosted under ``.pt'' \cite{silva-mario-03} can be used as reference. The publicly available legal documents and Wikipedia information in Tetun will also be collected. After gathering and building the collection of documents, it will be characterized according to each retrieval task.

\paragraph{Retrieval System Development}
Develop document and query processing methods for Tetun, such as tokenizer, spelling correction, stopwords removal, and stemmer. Afterward, develop the indexing and retrieval strategies and experiment with available indexing and ranking tools, including the retrieval and ranking models.

\paragraph{Building Test Collection}
Build the TREC-style test collection for evaluation based on the Cranfield approach \cite{clough-sanderson-13}. The relevance judgments will require Timorese expertise in different subjects, such as linguistics and computer science, and different backgrounds, mainly from academia and industry, to compare their assessment results. 

\paragraph{Evaluation}
Asses retrieval system effectiveness using different evaluation measures and then compare their results. Finally, identify the most effective solution that will be used to integrate with the Timor News platform for further evaluation.

\paragraph{Prototype Implementation}
Implement the prototypes in the Timor News production environment to evaluate the algorithms' performance and the interaction mechanism with the user using A/B tests.

\section{Research Question}
\label{sec:research_question}

This research is focused on developing solutions for text information retrieval in Tetun, therefore our main research questions are the following.

\paragraph{Research Question 1}
\textbf{\textit{What text pre-processing techniques can influence the text retrieval results in Tetun?}}
According to the textbook of Baeza-Yates and Ribeiro-Neto \cite[p.~8]{yates-neto-11}, an information retrieval system process comprise text transformation for both document and query, query expansion and modification, document indexing, retrieval, and ranking. The text transformation operations such as removing stopwords and stemming are also applied for both document and query. For queries, operations such as spelling correction and synonym expansion are also enforced. Considering that text pre-processing methods for Tetun do not exist and will be developed, this research question can extend our knowledge of those influencing the text retrieval results for a given query written in Tetun.
 
\paragraph{Research Question 2}
\textbf{\textit{What indexing and retrieval strategies provide the most effective solutions for Tetun text search?}}
Text indexing is a central part of a retrieval system. Different indexing tasks might have different results. Although our preliminary experiment indicates that indexing the news title provides the most relevant documents to the given queries, and is supported by the textbook of Baeza-Yates and Ribeiro-Neto \cite[p.~30]{yates-neto-11}, more experiments are needed to prove this evidence. Moreover, since in recent years several works have reported the promising results of applying neural language models in ad-hoc retrieval tasks \cite{NogueiraJPL20, YangZL19, DaiC19, Lin19}, both traditional and neural-based approaches will be adopted in the experiments.
 
\paragraph{Research Question 3} \textbf{\textit{Which query processing techniques improve the quality of the documents retrieved?}}
Since query processing involves parsing, expansion, and modification steps, it is vital to understand in what condition it contributes to improving the retrieval results. This research question broadens our knowledge to analyze the retrieval solutions specifically to ensure that the search solutions developed are explainable and understandable.

\section{Preliminary Study Results}
\label{sec:preliminary_study_result}

We report a preliminary result related to the experiment conducted on ad-hoc retrieval in Tetun. We employed 442 documents and 5 search queries (or \textit{topics}) written in Tetun from the Timor News database. Documents are the news articles, and queries are the search logs captured by the platform. The retrieval system is developed using Apache Solr \cite{solr} as the indexing, retrieval, and ranking platform. 

As an initial experiment, we manually developed a list of stopwords for Tetun by translating the English stopwords available in the NLTK\footnote{\url{https://www.nltk.org/}} library into Tetun. Some of the stopwords are in \autoref{tab:stopwords}. Moreover, we developed a light rule-based stemmer for Tetun by stemming to the root the noun words terminated with \textbf{``saun''}, \textbf{``mentu''}, \textbf{``dór'}', and \textbf{``-teen''}, e.g., \textit{selebrasaun} (\textbf{selebra}: celebrate, selebrasaun: celebration), \textit{juramentu} (\textbf{jura}: promise, juramentu: promise), lohidór (\textbf{lohi}: lie, \textit{lohidór}: lier), and \textit{nauktén} (\textbf{nauk}: steal, nauk-teen: thief). For the text normalization, we developed a rule-based list, e.g., junho, junu $\rightarrow$ \textbf{juñu} (june), to correct the misspelled Tetun terms. The same rule-based list is also used in constructing the lists of abbreviation and synonym expansions. The misspelled, abbreviation expansion and synonym lists were built based on the information extracted from the Timor News search query logs.

\begin{longtable}{@{}lllll@{}}
\caption{Tetun Stopwords Translated from English.}
\label{tab:stopwords}\\
\toprule
\textbf{Tetun}       & English         & \textbf{Tetun} & English                   \\
\midrule
 no & and & husi & from \\
 atu & so that & ha'u & I \\
 ne'e & this & sira & they \\
 ne'ebé & which & tanba & because \\
 \bottomrule
\end{longtable}

For the experiments, we set up three configurations (default, with stopwords, and without stemming) for query and document processing (referred to \autoref{tab:config}), and four indexing for documents. The elements of the documents indexed and tested are the combination of (i)~title (ii)~title and content, (iii)~lead and content, and (iv)~title, lead, and content. Thus, we have 12 different retrieval strategies (\textit{runs}). For each of the 12 retrieval strategies and 5 queries, we produce top-30 ranked documents, therefore, we have a pool of 1,800 documents (\textit{results}) to be evaluated.

\begin{longtable}{@{}lccc@{}}
\caption{Configuration of the Document and Query Processing.}
\label{tab:config}\\
\toprule
                                                 & \multicolumn{1}{l}{Default} & \multicolumn{1}{l}{With stopwords} & \multicolumn{1}{l}{Without stemming} \\* \midrule
\endfirsthead
\endhead
\multicolumn{1}{l}{HTML removal}               & x                            & x                                   & x                                     \\* 
\multicolumn{1}{l}{Case folding (lowercasing)} & x                            & x                                   & x                                     \\* 
\multicolumn{1}{l}{Tokenization}               & x                            & x                                   & x                                     \\* 
\multicolumn{1}{l}{Text normalization}         & x                            & x                                   & x                                     \\* 
\multicolumn{1}{l}{Stopwords removal}          & x                            & \multicolumn{1}{l}{}               & x                                     \\* 
\multicolumn{1}{l}{Abbreviation expansion}     & x                            & x                                   & x                                     \\* 
\multicolumn{1}{l}{Synonym expansion}          & x                            & x                                   & x                                     \\* 
\multicolumn{1}{l}{Light stemming}             & x                            & x                                   & \multicolumn{1}{l}{}                 \\* \bottomrule
\end{longtable}

To build the collection for evaluation, the aforementioned retrieval approach is used, but the total number of ranked documents was reduced to ten (10), then we have 120 documents for all retrieval strategies. We analyze and merge 10 documents out of 120 based on their ranked scores in each retrieval strategy to be judged.

We developed relevance judgment guidelines for the evaluators to manually judge the relevance of the documents for the given query. We adopt the relevance scoring scale of TREC 2020 for the news track \cite{soboroff-et-al-2020-trec}, with scores ranging from 0 to 3 as in the following:

\begin{itemize}
    \item[] \textbf{0} - the retrieved document provides \textit{no useful information} (not relevant).
    \item[] \textbf{1} - the retrieved document provides \textit{some useful information} or contextual information that would help the user understand the broader story context of the query article (less relevant).
    \item[] \textbf{2} - the retrieved document provides \textit{significant information} (relevant).
    \item[] \textbf{3} - the retrieved document provides \textit{essential information} (highly relevant).
\end{itemize}

Five queries and ten relevant documents for each of the queries are incorporated into the guidelines and then distributed to five independent Timoreses evaluators. The evaluators judge the relevance of the documents for each query according to the relevance scoring scale. After evaluating, we compile the judgment results and use majority voting to select the final score for the documents retrieved for each query. Then, we produce 50 lists of relevant documents (or \textit{qrels}) used to evaluate the retrieval system’s effectiveness.

We use the latest version of \textit{trec\_eval}\footnote{https://trec.nist.gov/trec\_eval/} to evaluate our retrieval system’s effectiveness. We perform 12 \textit{runs} based on the 12 retrieval strategies. For each \textit{run}, we have 150 documents (5 queries x 30 documents). We format \textit{results} and \textit{qrels} according to the TREC format, and then use precision (P) at 5 and 10 (P@5, P@10), mean average precision (MAP), and normalized discounted cumulative gain (nDCG) evaluation metrics to evaluate the system's effectiveness. Results show that, overall, \textit{without stemming with title indexing} outperformed all the other retrieval strategies, achieving the nDCG of 0.79 and MAP of 0.73 (see \autoref{tab:evaluation}).

\begin{longtable}{@{}llllll@{}}
\caption{The Evaluation Results: Title (T), Lead (L), and Content (C).}
\label{tab:evaluation}\\
\toprule
\multicolumn{2}{l}{}                                                    & \textbf{MAP}               & \textbf{P@5}               & \textbf{P@10}              & \textbf{nDCG}  \\* \midrule
\endfirsthead
\endhead
\multicolumn{1}{l}{\multirow{4}{*}{\textbf{Default}}}          & T     & 0.709                      & \textbf{0.800}             & 0.533                      & 0.775          \\* 
\multicolumn{1}{l}{}                                           & T+C   & 0.662                      & 0.667                      & \textbf{0.567}             & 0.715          \\* 
\multicolumn{1}{l}{}                                           & L+C   & 0.649                      & 0.633                      & \textbf{0.533}             & \textbf{0.780}          \\* 
\multicolumn{1}{l}{}                                           & T+L+C & \textbf{0.652 }                     & 0.633                      & 0.517                      & 0.732          \\* \midrule
\multicolumn{1}{l}{\multirow{4}{*}{\textbf{With stopwords}}}   & T     & \multicolumn{1}{l}{0.572} & \multicolumn{1}{l}{0.600} & \multicolumn{1}{l}{0.483} & 0.715          \\* 
\multicolumn{1}{l}{}                                           & T+C   & 0.612                      & 0.600                      & 0.533                      & 0.705          \\* 
\multicolumn{1}{l}{}                                           & L+C   & 0.593                      & 0.600                      & 0.500                      & 0.688          \\* 
\multicolumn{1}{l}{}                                           & T+L+C & 0.648                      & 0.667                      & 0.517                      & 0.729          \\* \midrule
\multicolumn{1}{l}{\multirow{4}{*}{\textbf{Without Stemming}}} & T     & \textbf{0.731}             & \textbf{0.800}             & \textbf{0.550}             & \textbf{0.790} \\* 
\multicolumn{1}{l}{}                                           & T+C   & \textbf{0.671}             & \textbf{0.700}             & 0.550                      & \textbf{0.730} \\* 
\multicolumn{1}{l}{}                                           & L+C   & \textbf{0.650}             & \textbf{0.700}             & 0.517                      & 0.719 \\*
\multicolumn{1}{l}{}                                           & T+L+C & 0.650             & \textbf{0.700}             & \textbf{0.550}             & \textbf{0.738} \\* \bottomrule
\end{longtable}

According to Baeza-Yates and Ribeiro-Neto \cite[p~29-30]{yates-neto-11}, the page title is usually shown prominently in web search, and a page title and combination of the title containing query terms are one of the search attributes that leads a search result to more clicks. This evidence reflects our result, the effectiveness of indexing documents by the title. However, since Timor News has established its news writing standard and designed it for the readers to understand the news content quickly from the title, we believe that it also contributes to the retrieval effectiveness result. Although without stemming terms to the root achieved the best result, since the overall scores of the default configuration (with stemming) is slightly lower, we assume that it can be affected by the quality of stemming rules and the amount of the documents and queries used in the experimentation. We take it into account to be investigated further in the future.

\section{Conclusion and Future Work}
\label{sec:conclusion_future_work}

This paper presents our preliminary results from the first ad-hoc retrieval experience using documents and search queries in Tetun collected from the Timor News database. We combined different indexing and retrieval strategies in the experiment to study the impact of each strategy on the quality of retrieved documents for each query. The experiment result provides a preliminary answer to the research questions~1 and~2 in Section \ref{sec:research_question}. For future work, we will follow the steps for developing text information retrieval for Tetun outlined in Section \ref{sec:methodology}.
\section{Acknowledgment}
This work is financed by National Funds through the Portuguese funding agency, FCT - Funda\c{c}\~{a}o para a Ci\^{e}ncia e a Tecnologia under the scholarship grant reference number SFRH/BD/151437/2021.

\bibliography{fdia}

\appendix

\end{document}